\documentclass[10pt,twocolumn,a4paper]{article}

\usepackage[left=20mm, right=15mm, top=15mm, bottom=15mm]{geometry}

\usepackage{subfig}
\usepackage{flushend}
\usepackage{indentfirst}
\usepackage{graphics}
\usepackage{amsmath}
\usepackage{graphicx}
\usepackage{epstopdf}
\usepackage{float}
\begin{document}

\title{\huge \textbf{Homotopy Perturbation Method for Solving
a Spatially Flat FRW Cosmological Model
}}

\date{}

\twocolumn[
\begin{@twocolumnfalse}
\maketitle

\author{\textbf{Victor Shchigolev}$^{1,*}$\\\\
\footnotesize $^{1}${Department of Theoretical Physics, Ulyanovsk State University, 42 L. Tolstoy Str., Ulyanovsk 432000, Russia}\\

\footnotesize $^{*}$Corresponding Author: vkshch@yahoo.com}\\\\\\

\end{@twocolumnfalse}
]

\noindent \textbf{\large{Abstract}} \hspace{2pt} In the present paper, we study a homogeneous cosmological model in Friedmann-Robertson-Walker (FRW) space-time by means of the so-called Homotopy Perturbation Method (HPM). First, we briefly recall the main equations of the cosmological model and  the basic idea of HPM. Next we consider the test example when the exact solution of the model is known, in order to approbate the HPM in cosmology and present the main steps in solving by this method. Finally, we obtain a solution for the spatially flat FRW model of the universe filled with the dust and quintessence when the exact solution cannot be found. A comparison of our solution with the corresponding numerical solution shows that it is of a high degree of accuracy.\\

\noindent \textbf{\large{Keywords}} \hspace{2pt} Cosmological Model, Exact and Approximate Solutions, Homotopy Perturbation Method \\

\noindent \textbf{\large{PACS }} \hspace{2pt}   98.80.-k; 98.80.Jk; 04.20.Jb, 04.25.-g, 02.30.Mv \\

\noindent\hrulefill

\section{\Large{Introduction}}

The advantages of analytical solutions in cosmology is hard to overestimate.  At the same time, obtaining exact solutions is very often problematic, and not always possible. This is related to the fundamental non-linearity of the basic equations in cosmology and gravitation. This problem becomes particularly difficult in the presence of various sources of gravity, especially given their nonlinear nature. At these circumstances, various approximate methods of solving can be used, such as the weak-field approximation in General Relativity \cite{Landau,Sushkov}, the slow-roll approximation in inflationary cosmology \cite{Guth,Linde}  etc. Noteworthy that in the course of such approximations, one has to ignore some terms in the equations, thereby losing the universality of the obtained solutions. The basic equation governing dynamics of the cosmic evolution  is known as the Friedmann equation. Since this equation is relevant to many cosmological models, any approach to its solving is always of great interest (see, for example, \cite{Copeland} and references therein).

The homotopy perturbation method was first proposed by Dr. He \cite{He1} in 1999 for solving differential and integral equations. This method has been extensively studied over a number of years  and successfully developed by numerous authors (e.g.  [7-26] to mention only a few). As well known, this method is a combination of homotopy in topology and classic perturbation techniques. The HPM has a significant advantage in that it provides an analytical approximate solution to a wide range of nonlinear problems in applied sciences.

The applications of the HPM cover the nonlinear differential equations, nonlinear integral equations, nonlinear differential-integral equations, difference-differential equations, and fractional differential equations. It has been shown that this method allows us to solve effectively, easily, and accurately a large class of nonlinear problems, and generally one or two iterations can lead to highly accurate solutions. The HPM yields a very rapid convergence of the solution series in most cases considered so far in literature. In the present paper, we show that the  usage of this method in cosmology also can  provide good results where obtaining the exact analytical solutions is basically impossible.

\section{\Large{FRW Cosmological Model}}

The Einstein's field equations with a constant $\Lambda$ - term  can be written as
\begin{equation}\label{1}
R_{ik}- \frac{1}{2} g_{ik} R - \Lambda g_{ik} = T_{ik},
\end{equation}
where we assume  that the gravity coupling constant $8\pi G=1$ for the sake of simplicity. All other symbols have their usual meanings in the Riemannian geometry. Considering the matter as a perfect fluid with the energy density $\rho_m$ and pressure $p_m$, we have the following  tensor of energy-momentum (TEM) of matter
\begin{equation}\label{2}
T_{ik}= (\rho_m +p_m)u_i u_k -p_m\, g_{ik},
\end{equation}
where  $u_i = (1,0,0,0)$ is  4-velocity of the co-moving observer, satisfying $u_i u^i = 1$.
 The line element of a spatially flat FRW space-time can be represented by
$$
ds^2 = d t^2- a^2 (t)\Big( dx^2+dy^2+dz^2\Big),
$$
where $a(t)$ is a scale factor of the Universe.
Given this metric and equation (\ref{2}), we can reduce  (\ref{1}) to the following set of equations:
\begin{eqnarray}
3H^2 &=& \rho_m +\Lambda,\label{3}\\
2 \dot H + 3H^2 &=& -  p_m+\Lambda,\label{4}
\end{eqnarray}
where $H = \dot a/a $ is the Hubble parameter, and the overdot stands for  differentiation with respect to cosmic time $t$.

The continuity equation follows from (\ref{3}) and (\ref{4}) as:
\begin{equation}\label{5}
\dot \rho_m + 3 H \Big(\rho_m + p_m \Big)=0.
\end{equation}

One could readily  verify that this continuity equation is equivalent to the conservation equation for TEM  $T^k_{i\,; k} = 0$, where semicolon stands for the covariant derivative, which is a direct consequence of the identity  $G^k_{i\,; k} = 0$ for the Einstein tensor.

Integrating the continuity equation (\ref{5}) for a constant equation of state (EoS) parameter $w_m=p_m/\rho_m$, we have
\begin{equation}\label{6}
\rho_m = \rho_{m0} \,a^{-3(1+w_m)},
\end{equation}
where $\rho_{m0}$ is a constant of integration. By substitution of (\ref{6}) into the Friedmann equation (\ref{3}) along with the Hubble parameter, we obtain the following main equation of the model
\begin{equation}\label{7}
\dot a^2 = H_{\Lambda}^2\Big[ \Omega_{m\Lambda}\, a^{-(1+3 w_m)} + a^2\Big],
\end{equation}
where
\begin{equation}\label{8}
H_{\Lambda} = \sqrt{\frac{\Lambda}{3}},\,\,\,\Omega_{m\Lambda}=\frac{\rho_{m0}}{\rho_{\Lambda}}=\frac{\rho_{m0}}{3H_{\Lambda}^2}
\end{equation}
is the dimensionless density parameter of matter, and  $\rho_{\Lambda}=\Lambda/3$ is the vacuum energy density. By introducing the dimensionless cosmic time $\tau = H_{\Lambda} t$, we can rewrite the Frienmann equation (\ref{7}) as
\begin{equation}\label{9}
a '^2 =\Omega_{m\Lambda}\, a^{-(1+3 w_m)} + a^2,
\end{equation}
where the prime stands for the derivative with respect to $\tau$. The Friedmann equation (\ref{10}) is substantially non-linear, with the exception of the obvious case of quasi-vacuum EoS,  $w_m = -1$. So hereinafter we are interested in the EoS of matter $w_m \neq -1$.

\section{\Large{Basic Idea of HPM}}

In order to briefly recall the Homotopy Perturbation Method, let us consider the following nonlinear differential equation:
\begin{equation} \label{10}
A(u)=f(r), \,\,\, r \in \it{\Omega},
\end{equation}
supplied with boundary conditions $B(u, \partial u/\partial n)= 0;\, r \in \it{\Gamma}$,
where $A$ is a general differential operator, $B$ is a boundary operator, $f(r)$ is a known analytic function, $\it{\Gamma}$ is the boundary of the domain $\it{\Omega}$. Suppose the operator $A$ can be divided into two parts: $M$ and $N$. Therefore, (\ref{10}) can be rewritten as follows:
\begin{equation} \label{11}
M(u) +  N(u)= f(r).
\end{equation}
The homotopy $v(r, p): {\it{\Omega}}\times [0,1] \to {\it I\!\!R} $ constructed as follows \cite{He1}
\begin{equation} \label{12}
H(v, p)=(1- p) [M(v)- M(y_0)]+p\,[A(v)- f(r)] = 0,
\end{equation}
where $r \in {\it{\Omega}}$  and $p \in [0, 1]$ is an imbedding parameter, and $y_0$ is an initial approximation of (\ref{10}). Hence, one can see that
\begin{equation} \label{13}
H(v, 0)= M(v)- M(y_0)=0,
H(v, 1)=A(v)- f(r) = 0,
\end{equation}
and changing the variation of $p$ from $0$ to $1$ is the same as changing $H(v, p)$ from $M(v)- M(y_0)$ to $A(v)- f(r)$, which are called homotopic. In topology,
this is called deformation. Due to the fact that $0 \leq p \leq 1$ can be
considered as a small parameter, by applying the perturbation procedure, one can assume that the solution of (\ref{12}) can be expressed as a series in $p$, as follows:
\begin{equation}\label{14}
v=v_0+p v_1+p^2 v_2 + p^3 v_3 + ...\, .
\end{equation}
When we put $p \to 1$, then equation (\ref{12}) corresponds to (\ref{11}), and (\ref{14}) becomes the approximate solution of (\ref{11}), that is
\begin{equation}\label{15}
u(x)= \lim_{p \to 1} v =v_0+v_1+ v_2 +  v_3 + ...\, .
\end{equation}

It should be noted that the series (\ref{15} is convergent for most cases. However, the convergent rate depends upon the nonlinear operator $A(v)$. Sometimes, even the first approximation is sufficient to obtain the exact solution \cite{He1}. As it is emphasized in \cite{He1} and \cite{Cveticanin}, the second derivative of $N(v)$ with respect to $v$ must be small, because the parameter $p$ may be relatively large, i.e. $p \to 1$, and the norm of $L^{-1} \partial N/\partial v$ must be smaller than one, in order that the series converges.

\section{\Large{Test Example}}

Let us consider the case of FRW cosmology with the only form of matter represented by the pressureless dust. The EoS of such a matter is known as $w_m =0$. Therefore, the Friedmann equation (\ref{9}) after multiplying by the scale factor $a$ becomes as follows
\begin{equation}\label{16}
a\,a '\,^2 =\Omega_{m\Lambda} + a^3.
\end{equation}
The exact solution to this equation can be readily found in the form
\begin{equation}\label{17}
a_{exact}(\tau) = \left[\Omega_{m\Lambda}\cdot\,\sinh ^2\Big(\frac{3}{2}\tau\Big)\right]^{1/3}.
\end{equation}

Now on, we consider the HPM of solving equation (\ref{16}). For this end, we suppose the following homotopy
\begin{equation}\label{18}
a\,a '\,^2  - a^3 -p\,\Omega_{m\Lambda} = 0,\,\,p \in [0,1],
\end{equation}
and assume that the solution of (\ref{16}) can be expressed as a series in $p$ by
\begin{equation}\label{19}
a(\tau) = a_0(\tau) + p\, a_1 (\tau) + p^2 a_2(\tau) + ...\, .
\end{equation}
The initial condition $a_0(0)$ can be freely chosen. Here we set $a_0(0)=-a_1(0) =\tilde{a}_{0}=const.\neq 0$ and $a_i(0)=0$, where $i > 1$.
The substitution of (\ref{19}) into equation (\ref{18}) yields
\begin{eqnarray}
p^0&:&a_0\, a\,'_0\,^2 -  a_0^3 = 0,   \label{20}\\
p^1&:&2a_0\, a\,'_0\, a\,'_1+(a\,'_0\,^2 - 3 a_0^2)\,a_1-\Omega_{m\Lambda}=0,\label{21}\\
p^2&:&2a_0\, a\,'_0\, a\,'_2 +(a\,'_0\,^2 - 3 a_0^2)\,a_2+a_0\, a\,'_1\,^2 \nonumber\\
& &-3a_0\, a_1^2 +2a_1\, a\,'_0\, a\,'_1 =0, \label{22}\\
& &.\,\,.\,\,.\,\,.\,\,.\,\,.\,\,.\,\,.\,\,.\,\,.\,\,.\,\,. \nonumber
\end{eqnarray}
It is noteworthy that we obtain the set of linear equations. Their solutions with the initial conditions given above can be readily found as
\begin{eqnarray}
\label{23}  a_0(\tau) &=& \tilde{a}_{0}\, e^{\tau}, \\
\label{24}  a_1(\tau) &=& \tilde{a}_{0}\, e^{\displaystyle\tau}\Big[\tilde{\Omega}_{m}(1-e^{-3\tau})-1\Big], \\
 a_2(\tau) &=& \tilde{a}_{0}\, e^{\tau}\Big[\frac{7}{6}\tilde{\Omega}_{m} (1-\tilde{\Omega}_{m})(1- e^{-3\tau}) \nonumber \\
\label{25} & &
+\frac{1}{4}\tilde{\Omega}_{m}^2(1-e^{-6\tau})\Big],
\end{eqnarray}
where
\begin{equation}\label{26}
\tilde{\Omega}_{m}=\frac{\Omega_{m\Lambda}}{6 \tilde{a}_{0}^3}.
\end{equation}

In accordance with the HPM, it follows  from (\ref{15}) and (\ref{23}) - (\ref{25}) that the solution of equation (\ref{16}) is given by
\begin{eqnarray}\label{27}
a(\tau)=\frac{\tilde{a}_{0}}{12}\, e^{ \tau}\Big[\Big(26-11\tilde{\Omega}_{m}\Big)\tilde{\Omega}_{m}-\nonumber\\
-2\Big(13-7\tilde{\Omega}_{m}\Big)\tilde{\Omega}_{m}\,e^{-3\tau}-3 \tilde{\Omega}_{m}^2\,e^{-6\tau}-... \Big].
\end{eqnarray}
In order to determine the free parameter $\tilde{a}_{0}$ in this solution, let us compare  (\ref{27}) with a series expansion of the exact solution (\ref{17}) in $\exp(-2\tau)$ represented by
\begin{equation}\label{28}
a_{exact}(\tau) = \sqrt[3]{\,\frac{\Omega_{m\Lambda}}{4}}\,e^{\tau} \left[1-\frac{2}{3}\,e^{-3\tau}-\frac{1}{9}\,e^{-6\tau}-... \right].
\end{equation}
If we put, for example,  $\Omega_{m\Lambda}=4,\Rightarrow\, \tilde{\Omega}_{m}=2/3\tilde{a}_{0}^3$ for the sake of simplicity, then we can rewrite (\ref{27}) as follows
\begin{eqnarray}\label{29}
a(\tau)=\tilde{a}_{0}\, e^{ \tau}\Big[\frac{1}{18\tilde{a}_{0}^3}\Big(26-\frac{22}{3\tilde{a}_{0}^3}\Big)-\,\,\,\,\,
\,\,\,\,\,\,\,\,\,\nonumber\\
-\frac{1}{9\tilde{a}_{0}^3}\Big(13-\frac{14}{3\tilde{a}_{0}^3}\Big)\,e^{ -3\tau}-\frac{1}{9\tilde{a}_{0}^6}\,e^{-6\tau}-... \Big].
\end{eqnarray}

The behavior of this solution in time compared with the exact solution (\ref{17}) is shown in Fig. 1.
\begin{figure}[thbp]
\centering
\includegraphics[width=0.4\textwidth]{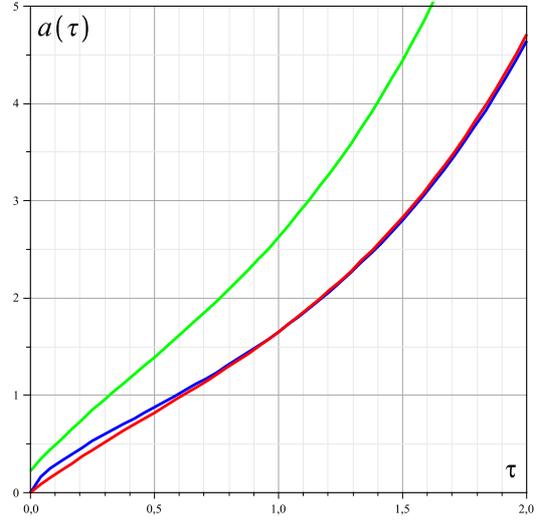}
\caption{The scale factor $a(\tau)$ is plotted against the dimensionless cosmic time $\tau$ for the exact solution (\ref{17}) (blue line), the approximation (\ref{28}) (green line), and for the HPM solution (\ref{29}) with $\tilde{a}_{0}=2.92$ (red line).}
\label{Figure_1}
\end{figure}
As can be seen, this solution approaches the exact solution (\ref{17}) in a much greater degree than the approximate solution (\ref{28}) of the same order. It should be emphasized that in the HMP solution we are able to fine-tune the model by means of the free parameter $\tilde{a}_{0}$. It is why we used $\tilde{a}_{0}=2.92$ when plotting $a(\tau)$ according to (\ref{29}) in Fig. 1.

Thus, we can see that a fairly high degree of accuracy is achieved in solving the Friedmann equation by the HPM,  even by means of two iterations.

\section{\Large{The Model with Dust and Quintessence}}

In this case, we have in equation (\ref{9}) one more term of the form (\ref{8}) for a quintessence with $w_m=w_q < -1/3$. If we consider  for example $w_q=-2/3$, then we can obtain the following Friedmann equation instead of (\ref{16})
\begin{equation}\label{30}
a\, a '\,^2 =\Omega_{m\Lambda} +\Omega_{q\Lambda}a^2 + a^3,
\end{equation}
where $\Omega_{m\Lambda}$ is defined by (\ref{8}), and $\Omega_{m\Lambda}=\rho_{q0}/3H_{\Lambda}$.

The Friedmann equation (\ref{16}) can be solved only in quadratures, and does not have an exact analytical solution.  So the attempt to solve this equation with the HPM and a comparison of it with the corresponding numerical solution are of certain interest.

Keeping in mind that we know the exact solution of (\ref{20}) given by (\ref{23}), now we suppose the following homotopy
\begin{equation}\label{31}
a\,a '\,^2  - a^3 -p\,(\,\Omega_{m\Lambda}+\Omega_{q\Lambda}\,a^2\,) = 0,\,\,p \in [0,1],
\end{equation}
and assume that the solution of (\ref{31}) can be expressed as a series in $p$ by (\ref{19}) as well. Here, we set $a_0(0)=\tilde{a}_{0}=const.\neq 0$ and $a_i(0)=0$, where $i\neq 0$.
The substitution of (\ref{19}) into equation (\ref{31}) yields
\begin{eqnarray}
p^0&:&a\,'_0\,^2 -  a_0^2 = 0,   \label{32}\\
p^1&:&2a_0\, a\,'_0\, a\,'_1+(a\,'_0\,^2 - 3 a_0^2)\,a_1\nonumber\\& & \,\,\,\,\,\,\,\,\,\,\,\,\,\,\,\,\,\,\,\,\,\,\,\,\,\,\,\,\,\,\,\,\,\,\,\,\,\,\,\,-\Omega_{m\Lambda} -\Omega_{q\Lambda}\,a_0^2 =0,\label{33}\\
p^2&:&2a_0\, a\,'_0\, a\,'_2 -2\, a_0^2\,a_2+a_0\,( a\,'_1\,^2 -3a_1^2)\nonumber\\
& &\,\,\,\,\,\,\,\,\,\,\,\,\,\,\,\,\,\, +2 a\,'_0\,a_1\, a\,'_1 - 2 \Omega_{q\Lambda}\,a_0\, a_1=0. \label{34}
\end{eqnarray}

One can readily verify  that equation (\ref{23}) represents the exact solution of (\ref{32}). Using this solution, one can obtain the exact solutions of (\ref{33}) and (\ref{34}) as follows
\begin{equation}\label{35}
a_1(\tau)=\tilde{a}_{0}\,e^{\tau}\left[\,\tilde{\Omega}_{q}\left(1-e^{-\tau}\right)+
\tilde{\Omega}_{m}\left(1-e^{-3\tau}\right)\right],
\end{equation}
and
\begin{equation}
a_2(\tau) = e^{\tau}\Big[-(2\tilde{a}_{0}\tilde{\Omega}_{q}-1)(\tilde{\Omega}_{q}+\tilde{\Omega}_{m})(1-e^{-\tau})+ \phantom{.......} \nonumber \\
\end{equation}
\begin{equation}
\frac{\tilde{\Omega}_{q}}{4}(3\tilde{a}_{0}\tilde{\Omega}_{q}\!-\!2)(1-e^{-2\tau}) -2\tilde{a}_{0}\tilde{\Omega}_{m}(\tilde{\Omega}_{q}+\tilde{\Omega}_{m})(1-e^{-3\tau})  \nonumber
\end{equation}
\begin{equation}
-\frac{\tilde{\Omega}_{m}}{4}(5\tilde{a}_{0}\tilde{\Omega}_{q}+1)(1-e^{-4\tau})+
\frac{\tilde{a}_{0}\tilde{\Omega}_{m}^2}{4}(1-e^{-6\tau})\Big],\label{36}
\end{equation}
where $\tilde{\Omega}_{m}$ is defined by (\ref{26}),
\begin{equation} \label{37}
\tilde{\Omega}_{q}=\frac{\Omega_{q\Lambda}}{2\tilde{a}_{0}},
\end{equation}
and the initial condition $a_1(0)=a_2(0)=0$ is applied. Due to  (\ref{15}), it follows from (\ref{23}) and  the set of equations (\ref{35}), (\ref{36}) that the solution of (\ref{30}) is given by
\begin{eqnarray}\label{38}
a(\tau)= e^{\tau}\Big[A_0+A_{-1}\, e^{-\tau}+A_{-2}\, e^{-2\tau}  +A_{-3}\, e^{-3\tau}\nonumber\\+A_{-4}\, e^{-4\tau}+A_{-6}\, e^{-6\tau}\Big],
\end{eqnarray}
where
\begin{equation} \label{39}
\left.
\begin{array}
[c]{rcl}
& &A_0= \tilde{a}_{0}(\tilde{\Omega}_{q}+\tilde{\Omega}_{m}+1)+ {\displaystyle \frac{(2\tilde{\Omega}_{q}+3\tilde{\Omega}_{m})}{4}} \\ \\
& &\,\,\,\,\,\,\,\,\,\,-{\displaystyle \frac{\tilde{a}_{0}}{4}}(5\tilde{\Omega}_{q}^2+21\tilde{\Omega}_{q}\tilde{\Omega}_{m}+7\tilde{\Omega}_{m}^2), \\ \\
& &A_{-1} = (2\tilde{a}_{0}\tilde{\Omega}_{q}-1)(\tilde{\Omega}_{q}+\tilde{\Omega}_{m})-\tilde{a}_{0}\tilde{\Omega}_{q}, \\ \\
& &A_{-2} = -{\displaystyle \frac{1}{4}}\tilde{\Omega}_{q}(3\tilde{a}_{0}\tilde{\Omega}_{q}-2), \\ \\
& &A_{-3} = \tilde{a}_{0}\tilde{\Omega}_{m}(2\tilde{\Omega}_{q}+2\tilde{\Omega}_{m}-1), \\ \\
& &A_{-4} = {\displaystyle \frac{1}{4}}\tilde{\Omega}_{m}(5\tilde{a}_{0}\tilde{\Omega}_{q}+1),\\ \\
& &A_{-6} = -{\displaystyle \frac{1}{4}}\tilde{a}_{0}\tilde{\Omega}_{m}^2.
\end{array}
\right\}
\end{equation}

\begin{figure}[thbp]
\centering
\includegraphics[width=0.4\textwidth]{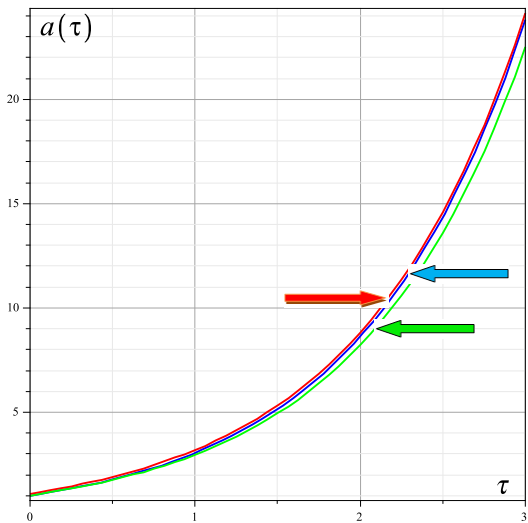}
\caption{The scale factor $a(\tau)$ is shown for the numerical solution of (\ref{30}) with $a(0)=1$  (blue line), obtained with the help of Maple, and for the HPM solution (\ref{38}) with $\tilde{a}_{0}=1.1$ (red line) and $\tilde{a}_{0}=1.0$ (green line).}
\label{Figure_2}
\end{figure}
As an illustrative example, we can set $\Omega_{m\Lambda}=0.4,\,\Omega_{q\Lambda}=1/35$. For these values of the relative densities of matter and, for example, $\tilde{a}_{0}=1.1$, one can find from (\ref{39}) that $A_0= 1.204623506,\,\, A_{-1}= -0.07555824735,\,\, A_{-2}= 0.006354359928, \,\,A_{-3}= -0.04814604228,\,\, A_{-4}= 0.01341633573,\,\, A_{-6}= -0.0006899125812$. The graphs of $a(\tau)$ for the numerical solution of  (\ref{30}) and the HPM solutions  (\ref{38}) with two different values of $\tilde{a}_{0}$ are plotted in Fig. 2. It can be seen that two lines which describe the numerical solution and the approximate solution (with $\tilde{a}_{0}=1.1$) look almost identical. It is quite remarkable that the further and better approximation could be achieved by means of the further fine-tuning of $\tilde{a}_{0}$.

Given the analytical solution (\ref{38}), we are able now to study this model further. It is not our aim here to consider all its features in detail. Note only that one can obtain almost all parameters of the model by means of the Hubble parameter $H = \dot a/a$. In view of (\ref{38}), it is easy to derive the following equation
\begin{equation}\label{40}
H(\tau)=H_{\Lambda}\frac{\displaystyle A_0- \sum_{n=2}^4 (n-1)A_{-n}\, e^{-n\tau}-5A_{-6}\, e^{-6\tau}}{\displaystyle A_0+\sum_{n=1}^4 A_{-n}\, e^{-n\tau}+A_{-6}\, e^{-6\tau}},
\end{equation}
where $H_{\Lambda}$ is defined by (\ref{8}). Hence we can see that $H(\tau \to \infty)=H_{\Lambda}$. In Fig. 3 we can observe the time behavior of the Hubble parameter in accordance to (\ref{40})  for two different values of  $\tilde{a}_{0}$ .

\begin{figure}[thbp]
\centering
\includegraphics[width=0.4\textwidth]{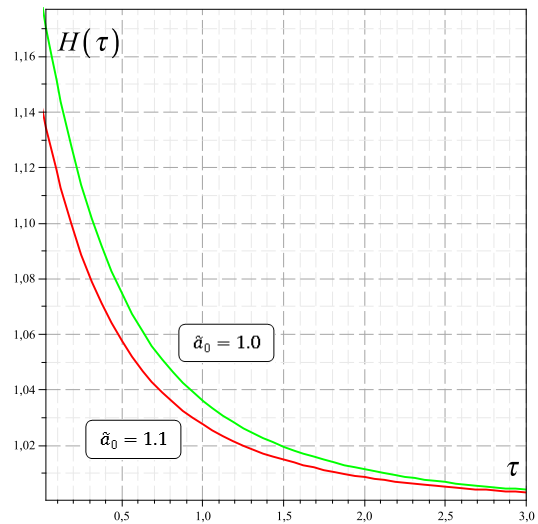}
\caption{The Hubble parameter $H(\tau)$ is plotted against the dimensionless cosmic time $\tau$ according to (\ref{40}) for $\tilde{a}_{0}=1.1$ (red line) and $\tilde{a}_{0}=1.0$ (green line). Here, we put $H_{\Lambda}=1$.}
\label{Figure_3}
\end{figure}

\section{\Large{Conclusions}}

Thus, we have studied a spatially flat FRW model with the help of the  Homotopy Perturbation Method.  We have considered the test example with the well-known exact solution  in order to approbate the HPM in the FRW cosmology and present the main steps in solving by this method. Moreover, we have obtained a solution for the spatially flat FRW model of the universe filled with the dust and quintessence when the exact solution could not be found. The comparison of our solution with the corresponding numerical solution showed the high degree of accuracy for the aproximate solution.

Of course, this method of solving the Friedmann cosmology is not universal. It also has several disadvantages inherent in any approximate method.   Unfortunately, this method is very sensitive to the choice of homotopy, which often determines the possibility of the rapid convergence of approximate solution to the exact one. Moreover,  it is not clear yet how to fine-tune the free parameter. We hope to consider all these  problems in our further works. In the preset paper, we just wanted to draw the attention of researchers to the possible application of the HPM in cosmology.

We did not aim to investigate the accuracy of the HPM and to estimate the errors of approximation. We simply showed that this method can be used with good results where it is impossible to solve the cosmological equations in explicit analytical form. Therefore,  this method in cosmology can be further developed in the context of even greater number of problems. In our view, the results of the present work reveal that the HPM is very effective and simple for obtaining approximate solutions of the Friedmann equation in cosmology.

\noindent\hrulefill

\end{document}